\documentclass[USenglish,twocolumn]{article}

\usepackage[utf8]{inputenc}
\usepackage{pdflscape}
\usepackage{epstopdf}
\usepackage[big]{dgruyter}

\begin{document}

  \articletype{Research Article{\hfill}Open Access}

  \author*[1]{K. I. Smirnova}

  \author[2]{D. S. Wiebe}
  
  \author[3,4]{A. V. Moiseev}

  \affil[1]{Ural Federal University, E-mail: Arashu@rambler.ru}
  
  \affil[2]{Institute of Astronomy, Russian Academy of Sciences, E-mail: dwiebe@inasan.ru}
  
  \affil[3]{Special Astrophysical Observatory, Russian Academy of Sciences, E-mail: moisav@sao.ru}
  
  \affil[4]{Space Research Institute, Russian Academy of Sciences}

  \title{\huge Star-forming complexes in the polar ring galaxy NGC660}

  \runningtitle{Star-forming complexes in the polar ring galaxy NGC660}


  \begin{abstract}
{Galaxies with polar rings consist of two subsystems, a disk and a ring, which rotate almost in orthogonal planes. In this paper, we analyze the parameters characterizing the composition of the interstellar medium and star formation in star-forming complexes, belonging to a polar ring galaxy NGC660. We show that star-forming regions in the ring of the galaxy are distinctively different from those in the galaxy disk. They possess substantially lower infrared luminosities, indicative of less dust mass in these regions than in a typical disk star-forming region. UV and H$\alpha$ luminosities also appear to be lower in the ring, probably, being a consequence of its relatively recent formation.}
\end{abstract}

  \journalname{Open Astronomy}
\DOI{DOI}
  \startpage{1}
  \received{..}
  \revised{..}
  \accepted{..}

  \journalyear{2014}
  \journalvolume{1}

\maketitle
\section{Introduction}

The interstellar medium (ISM) is a stage and a source of material for the star formation process. Its composition defines the rate and spatial distribution of star formation in a galaxy. So, in order to recover specific details of this process, we need to know the relative contribution of various ISM components (atomic gas, molecular gas, dust, radiation, etc.). Rapid development of observational tools allows studying these components separately from each other. Interacting galaxies are especially interesting from this point of view, where there exists a certain perturbing factor, triggering appearance of new star-forming regions (SFRs). This leads to a rejuvenation of the stellar population of the galaxy and may clear up some features of the ISM evolution.

Thus, we are interested in such objects, where we can observe young regions of star formation. Galaxies with polar rings are good candidates for this role. At the moment, about 400 candidates and confirmed polar ring galaxies are listed in two catalogs: \cite{1990Whitmore} and   \cite{2011Moiseev}.To perform a detailed analysis of observational tracers of various ISM components, we need a large set of observational data for each object, covering as wide wavelength range as possible. This condition is satisfied only in a few objects. For our research, we need an extended ring to isolate star-forming regions, belonging to it. We also need the ring to be well separated from the host galaxy so that we can analyze emission separately for the two subsystems. The galaxy should not be far away, so that individual star-forming complexes can be resolved.

So far, there seems to be just one galaxy suitable for our purposes. This is NGC660 galaxy, located at the distance of 13 Mpc \cite{2013Tully},  included in the catalogue \cite{1990Whitmore} under the number C-13. The galaxy NGC660 was confirmed to be a polar ring galaxy with a strongly inclined extended ring according to HI observation by \cite{1990Gottesman}. In contrast with most other polar rings, which surround early-type  gas-free galaxies, the central disk of NGC 660 contains molecular and atomic gas as well as star forming regions. Study of gas kinematics  provides evidence of direct interaction between matter in external ring and inner disk \cite{2014Moiseev}. 

In \cite{ourpreviouswork}, we investigated relations between various components of ISM in 11 galaxies, included in a number of observational surveys, including THINGS (HI line at 21 cm, Karl G. Jansky Very Large Array, VLA), KINGFISH (far infrared region, 70, 100, and 160 $\mu$m, Herschel Space Observatory), SINGS (NIR and MIR bands, 3.6, 4.5, 5.8, 8.0 and 24 $\mu$m, Spitzer Space Telescope), and HERACLES (CO 2–-1 line, IRAM). Infrared (IR) data were also used in \cite{ourpreviouswork} to estimate the total dust mass ($M$), the fractional mass of polycyclic aromatic hydrocarbons ($q_{\rm PAH}$), and the radiation field strength ($U$), using the model by Draine and Lee \cite{DL2007}. In this study, we want to take advantage of the obtained patterns for these `normal' galaxies and to compare them with the similar results for the galaxy NGC660. Data in the far IR range are not available for this galaxy, so we cannot use the Draine and Lee dust model. But we can deduce the radiation field that heats up the dust, using data in the ultraviolet range and the H$\alpha$ line.

\section{Data}

In this section we present the available observational data for the galaxy NGC660, as well as their processing. We have identified star-forming complexes in the disk and the ring of the galaxy. In all the ranges we performed the aperture photometry of the same complexes. A typical size of an aperture is about $15^{\prime\prime}$, which at the distance of NGC660 corresponds to a few hundred pc (the accepted scale is 63 pc$/\prime\prime$). Thus, the considered regions are actually large-scale star-forming complexes.

\subsection{Observations}

For the galaxy NGC660, we used archival data in the IR and UV ranges. Near-IR observations (wavelengths of 3.6, 4.5, 5.8 and 8.0 $\mu$m) are taken from the SINGS survey\footnote{http://sings.stsci.edu} performed on the Spitzer Space Telescope. The data in the mid-IR region (22 $\mu$m) were taken from the WISE archive\footnote{http://irsa.ipac.caltech.edu/frontpage/} \cite{2010Wright}, data in the ultraviolet range are extracted from the GALEX archive \cite{2005Martin}.
Observations in the H$\alpha$  emission line were performed with the 6-m telescope of the Special Astrophysical Observatory of the Russian Academy of Sciences (SAO RAS) using the scanning Fabry-Perot interferometer in the  multi-mode SCORPIO focal reducer \cite{2005Afanasiev}.

Since the point spread functions (PSFs) of different instruments and passbands differ from each other, we convolved near IR, mid IR and 21 cm images to the resolution of 22 $\mu$m WISE images, using the IDL convolution procedure and kernels provided by Aniano et al. \cite{2011Aniano}.

\subsection{Aperture Photometry}

\begin{figure}
\includegraphics[width=0.45\textwidth]{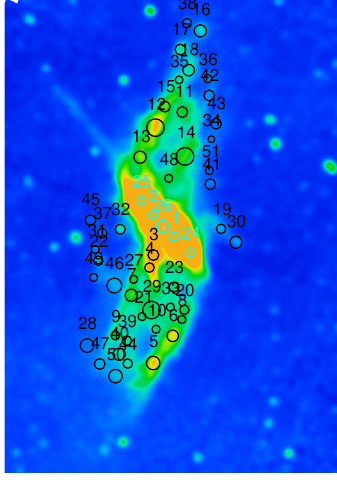}
\caption{An 8 $\mu$m image of NGC660 with overlayed regions. Regions 1, 2, 24, 25, 26 belong to the galaxy disk. Other regions are located in the ring. \label{fig:maps}}
\end{figure}

Regions for aperture photometry in the NGC660 galaxy were selected both in the disk of the galaxy and in its ring (Fig. \ref{fig:maps}). The ring of the galaxy has quite an extended structure, which allows selecting numerous star-forming complexes. The disk of the galaxy is seen almost edge-on, thus, we have only been able to identify few brightest regions there.


After the regions have been selected, the convolved images were used to perform aperture photometry. The aperture radius is chosen separately for each HII complex, depending on its angular size. The total flux was corrected for the contribution of partial pixels. Photometry procedure and error estimation are described in \cite{Khramtsova2013}. Eventually, we obtain radiation fluxes in the near and mid-IR bands as well as in the UV range and in H$\alpha$. A significant fraction of the emission at 8 and 22 $\mu$m is generated by stars. Also, some emission at 8 $\mu$m is produced  not only by  polycyclic aromatic hydrocarbons (PAHs), but also by hot larger dust grains. In order to distinguish only the PAH emission at 8 $\mu$m ($F_{8}^{\rm afe}$) and only the very small grains (VSG) emission at 22 $\mu$m ($F_{22}^{\rm ns}$), we applied the method for estimating the stellar and large grain contributions described in \cite{2010Marble}.

\section{Results}

The main goal of our study is the relation of dust and gas properties in the regions of star formation. We assume that radiation in the near IR, in particular, at 8 $\mu$m is generated by small aromatic particles of PAH, while emission in the mid-IR, in particular, at 22 and 24 $\mu$m is produced by larger hot dust grains. In both cases, to generate IR emission, it is necessary to heat dust particles by ultraviolet radiation. Therefore, we assume that spots of intense near- and mid-IR radiation are SFRs.

\begin{figure}
\includegraphics[width=0.45\textwidth]{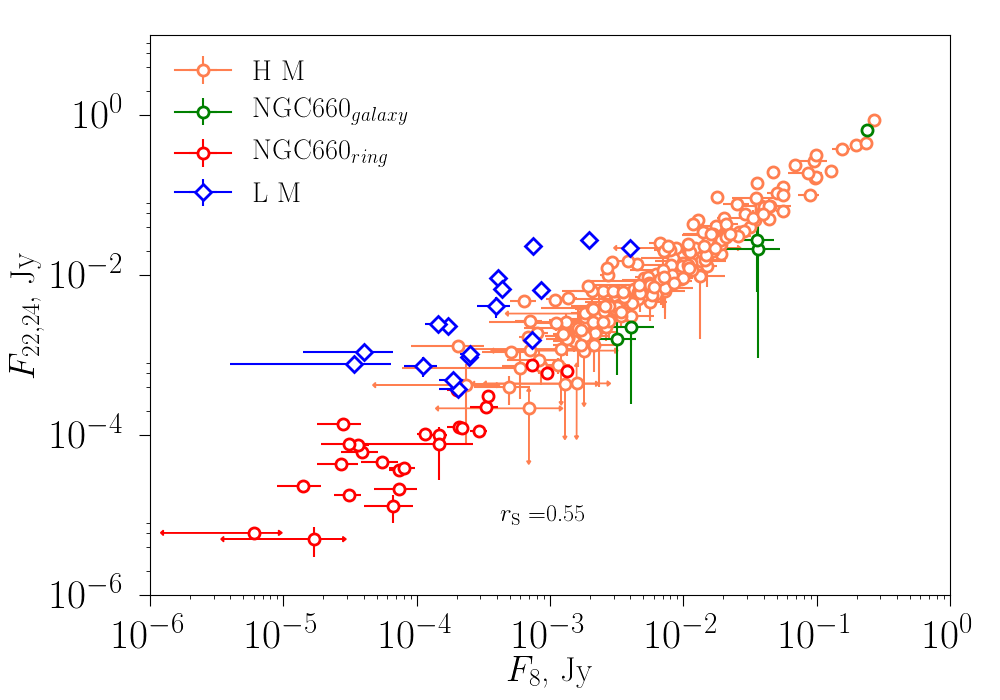}
\caption{Comparison of fluxes from SFRs in NGC660 at 8 $\mu$m and 22 $\mu$m. For comparison, 8 and 24 $\mu$m fluxes are also shown for high and low metallicity extragalactic SFRs in 11 galaxies studied in \cite{ourpreviouswork}.}
\label{ngc660_822}
\end{figure}

The general relation of fluxes at 8 and 22 $\mu$m is shown in Fig.~\ref{ngc660_822}. Also, in this graph we show fluxes at 8 and 24~$\mu$m from SFRs studied in \cite{ourpreviouswork}. Orange circles correspond to SFRs in galaxies of high metallicity. Blue diamonds show SFRs in galaxies of low metallicity. As noted in \cite{ourpreviouswork}, data points for low metallicity SFRs in this diagram are located slightly above the points for SFRs of high metallicity. This indicates a known dependence of flux ratio at 8 $\mu$m to 24 $\mu$m (or 22 $\mu$m) on metallicity \cite{engelbraht}. Clearly, the SFRs from the NGC660 disk (green circles) are located in the same region of the diagram as the high-metallicity SFRs from the other star-forming galaxies (the sample from \cite{ourpreviouswork} includes both spirals and irregulars). The SFRs from the NGC660 ring (red circles) occupy a different position in the diagram. While demonstrating the same correlation between the two fluxes as the other SFRs of high metallicity, they show substantially lower fluxes, both at 8 $\mu$m and at 22 $\mu$m.

\begin{figure}
\includegraphics[width=0.45\textwidth]{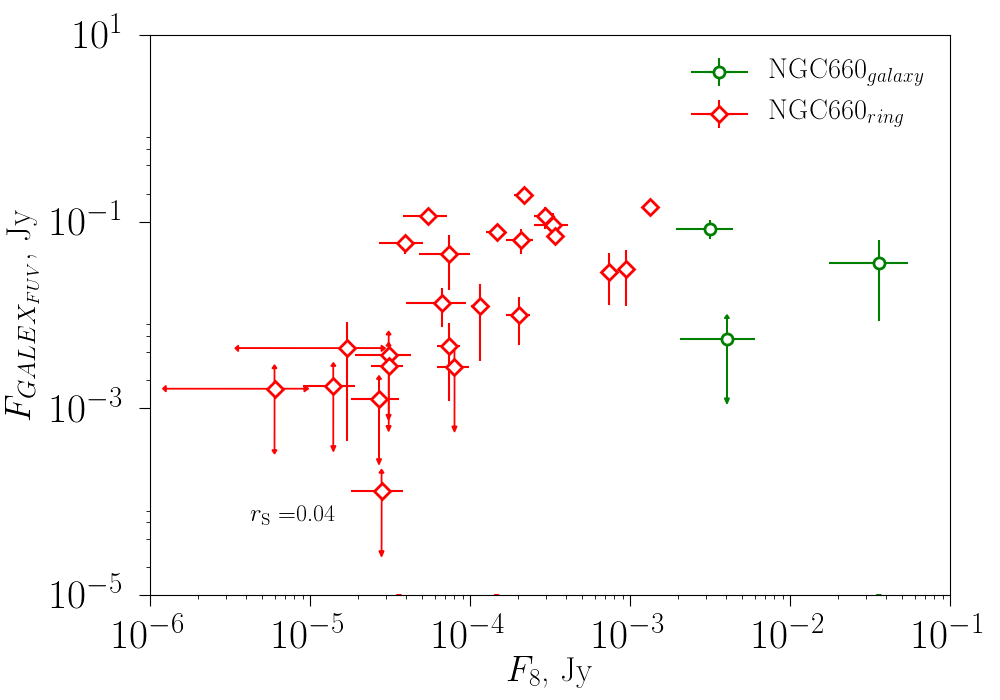}
\includegraphics[width=0.45\textwidth]{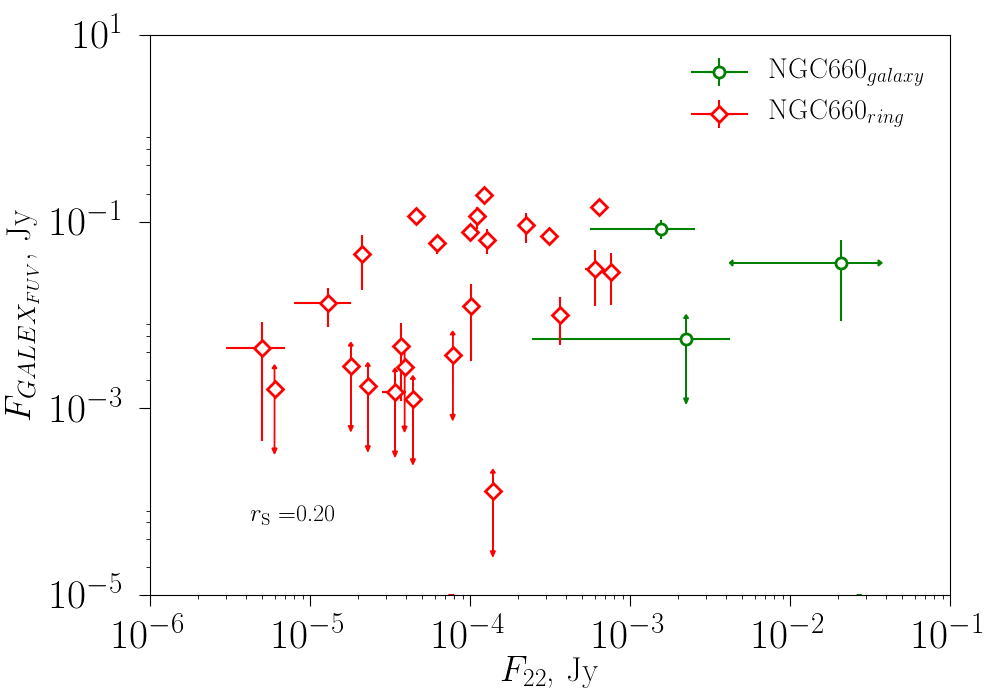}
\caption{Comparison of aperture photometry results for SFRs in the galaxy NGC660 at wavelengths of 8 and 22 $\mu$m and the GALEX FUV filter.}
\label{ngc660_UV}
\end{figure}

The low brightness of the IR radiation of the ring regions can be caused both by a small number of emitters (aromatic particles) and/or by a low intensity of exciting ultraviolet emission. The latter assumption can be qualitatively checked using GALEX images. To compare the emission in the infrared and in the UV range, in Fig. \ref{ngc660_UV} we compare results of aperture photometry for  Spitzer and GALEX observations in the FUV filter (data for the NUV filter show similar trends). In the top panel of Fig.~\ref{ngc660_UV} the results of ultraviolet observations are related to the 8 $\mu$m radiation. Obviously, the intensity of aromatic bands in the ring correlates well with the ultraviolet emission. The situation is less clear with the radiation from the SFRs of the galaxy disk. On the $F_8-F_{\rm FUV}$ diagram the disk SFRs are clearly separated from the ring SFRs. The 8 $\mu$m flux in the disk regions is nearly an order of magnitude greater than in the ring regions. At the same time, the disk SFRs are characterized by the similar or even lower UV fluxes. Note also that we have only succeeded in determining the UV flux in three of five disk regions. In the other two regions the UV emission merges with the background. The correlation of the 22 $\mu$m flux with the FUV filter flux is also visible even though it is weaker (bottom panel of Fig.~\ref{ngc660_UV}). And again we see the order of magnitude higher 22 $\mu$m flux in the disk SFRs coupled to the moderate UV luminosity.


\begin{figure}
\includegraphics[width=0.45\textwidth]{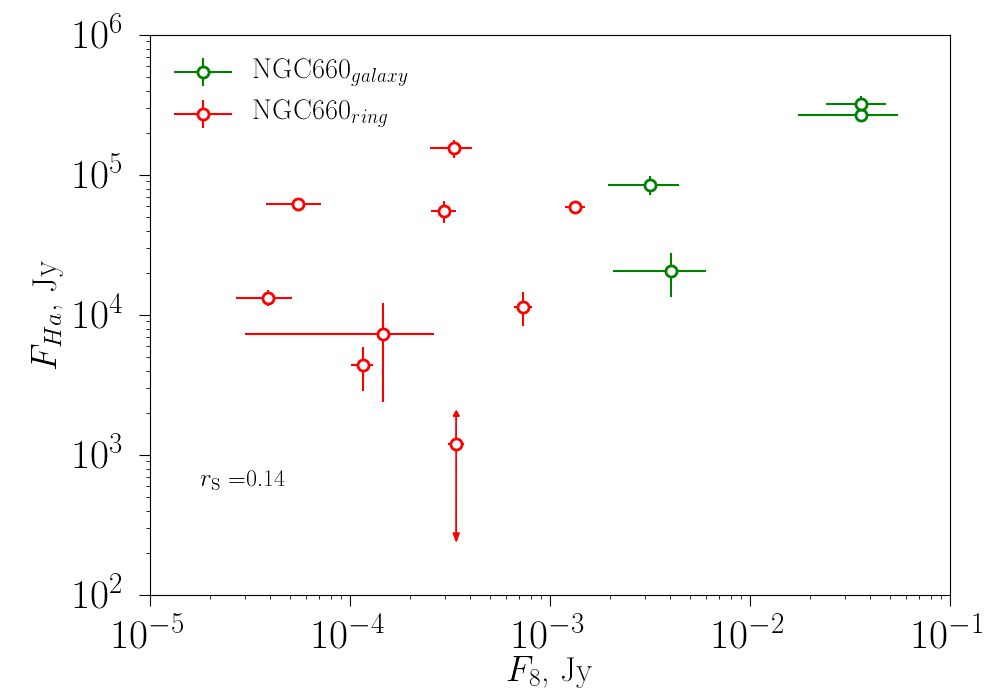}
\includegraphics[width=0.45\textwidth]{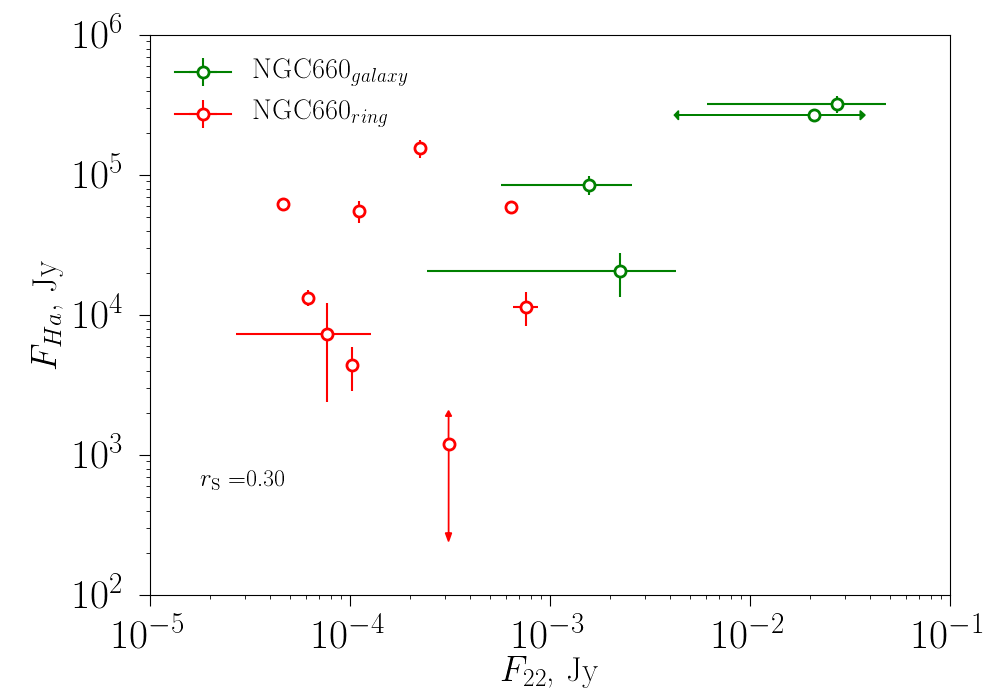}
\caption{Comparison of H$\alpha $ fluxes with the 8 $\mu$m (upper) and 24 $\mu$m (bottom) fluxes.}
\label{ngc660_Ha}
\end{figure}

The correlation between the infrared fluxes and the emission in the H$\alpha$ line is weaker, which is illustrated in Fig.~\ref{ngc660_Ha}. However, in this case the disk SFRs and the ring SFRs do seem to follow the same trend of having higher H$\alpha$ line intensities in regions with higher infrared luminosities. The disk SFRs are noticeably brighter both in H$\alpha$ and in the IR.

\section{Conclusions}

Our analysis shows that SFRs in the disk and in the ring of NGC660 appear to have quite different properties. The disk SFRs are similar in their emission characteristics to SFRs in `normal' disk and irregular galaxies studied in \cite{ourpreviouswork}. At the same time, the ring SFRs are more compact and show substantially fainter emission in the infrared bands. One may argue that this is a selection effect, as in the ring, which is a less crowded environment, we are able to see smaller and fainter regions than in a typical galactic disk. This is an important consideration, which deserves further study, however, we note that even in the dwarf irregular Holmberg~II, where SFRs are quite isolated from each other, they are larger and brighter that the SFRs in the NGC660 ring \cite{ourpreviouswork}.

The ring SFRs in NGC660 also appear to be less bright in H$\alpha$. The situation is unclear with the UV emission. The disk SFRs and the ring SFRs do overlap in the sense that the UV bright ring SFRs are comparable (or even exceed) in luminosity to the disk SFRs. This may be related to opacity effects. If we assume that the regions that are brighter in the infrared are richer in dust, they should be more opaque in the UV band, so that there is a certain `saturation' effect in their UV luminosity, but this suggestion needs to be verified on a larger SFR sample. We may also start to see this effect in H$\alpha$, where there does exist a difference between the disk SFR H$\alpha$ emission and the ring SFR H$\alpha$ emission, but this difference is much less prominent than in the infrared.

Overall, we conclude that SFRs in the ring of NGC660 are smaller than the disk SFRs and contain less dust, at least, they are less rich in those dust particles that emit in the near- and mid-IR bands. They are also characterized by fainter H$\alpha$ and UV luminosities, which may indicate their relative youth. In the future we plan to extend this study, attracting kinematical data on the SFRs in NGC660 as well as in other galaxies with recently triggered star formation.

DW thanks the OFN-17 grant for support. The work of KS was supported by Act 211 Government of the Russian Federation, contract № 02.A03.21.0006. This work is based in part on observations made with the Spitzer Space Telescope, which is operated by the Jet Propulsion Laboratory, California Institute of Technology under a contract with NASA. Data products from the Wide-field Infrared Survey Explorer, which is a joint project of the University of California, Los Angeles, and the Jet Propulsion Laboratory/California Institute of Technology, funded by the NASA, are utilized.

%

\end{document}